\documentclass[a4]{IEEEtran}

\usepackage{hyperref}
\usepackage{graphicx}

\title{$M^3$: An Open Model for Measuring Code Artifacts}

\author{ Anastasia Izmaylova, Paul Klint, Ashim Shahi and Jurgen Vinju}

\date{}

\newcommand{\loc}[1]{\small{\texttt{#1}}}
\newcommand{\mthree}{\ensuremath{M^3}}

\begin{document}

\maketitle

\section{Motivation}

In the context of the EU FP7 project ``OSSMETER'' we are developing an infra-structure for measuring source code. The goal of OSSMETER is to obtain insight in the quality of open-source projects from all possible perspectives, including product, process and community. 

The main challenge that our part of the design, which focuses on code, faces is variability: the different languages we support as well as the different metrics we will compute. The standard solution is to put an explicit model (database, graph) in between such that model producers (parsers \& extractors) can be de-coupled from model consumers (metrics \& visuals). This abstract is a ``white paper'' on \mthree, a set of code models, which should be easy to construct, easy to extend to include language specifics and easy to consume to produce metrics and other analyses. We solicit feedback on its usability.

\begin{figure}[t]
\includegraphics[width=\columnwidth]{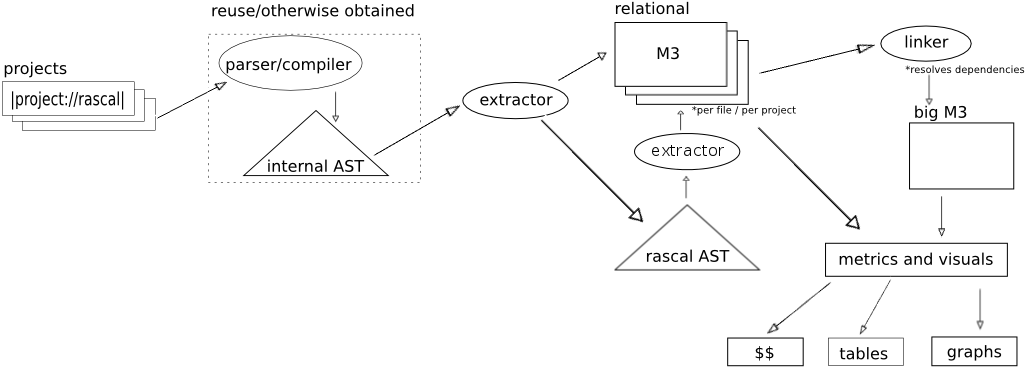}
\caption{An overview: from code to metrics and visuals via M3 models.}
\end{figure}

The context of \mthree\ is the Rascal meta-programming language\footnote{\url{http://www.rascal-mpl.org}}. This is a domain specific language specifically designed to include primitives we need to model any programming language syntax and semantics, and to analyze and manipulate these models. Three essential design elements for the purpose of this paper are that Rascal has value semantics for all in-memory data, including sets and relations, it has support for URI literals, called ``source locations'', and it has term rewriting and relational calculus primitives to deal with hierarchical and relational data, respectively. This includes generic traversal and pattern matching primitives as well as relational operators such as transitive closure and comprehensions. 

Caveat emptor. The reader should be aware that we do \emph{not} intend to create a unified model for programming language semantics. Such a language independent model would be inaccurate (wrong), and deliver meaningless metrics. Instead we opt for a unified \emph{form} for storing facts about programs. This means that all models will have a predictable shape, but we do not assume any reusability of metrics or visuals producers between models produced by different parsers. 

\mthree\ is inspired by models such as FAMIX, RSF, GXL, ATerms and S-Expressions. The differences are that \mthree\ deals with purely immutable, typed, data and can be directly produced, manipulated and analyzed using Rascal primitives. Two unique elements are the introduction of URI literals to identify source code artifacts in a language agnostic manner and support for fully structured type symbols. Otherwise \mthree\  is very similar in intent and solution patterns to the aforementioned existing models.

\section{Design aspects}

\subsection{Textual models}

M3 is, like all Rascal data, fully typed and fully serializable as readable text with a standard notation that is equal to the expression syntax for literals. This means that any intermediate step can be visualized as plain text and not only searched and edited using standard text editing facilities, but also stored and retrieved persistently. One particular aspect of the Rascal IDE is that all printed source location literals (see below) in editors and consoles are treated as \emph{hyperlinks}. M3 models are therefore ``programmer friendly'': easy to explore both inter-actively and programmatically using low-brow techniques.

\subsection{Locations} 

To verify the correctness of metrics or for explaining them we want to trace back measurements to code. For example, when we present the largest class in a project, we need the size as well as a link to the source code of this class. In other words, want to link information back to source code for all derived facts we produce. From the semantic web we take the idea of using URI (Uniform Resource Identifiers) to model the identity of any artifact. Each URI takes the following shape: \loc{|<scheme>://<auth>/<path>?<qry>|(<off>,<len>)}.

We distinguish between two kinds of code locations: physical and logical. A \emph{physical} location identifies a storage location. Physical locations may be absolute or relative. Examples of absolute physical locations are \loc{|file:///tmp/Hello.java|} and \loc{|http://foo.com/index.html|}, and \loc{|project://MyPrj/Hello.java|} is a relative location. It is always the \emph{scheme} of a URI that defines to which root a URI is relative. In the case of \texttt{project}, it is an Eclipse project in the current workspace, in the case of \texttt{cwd} it is the current working directory. The set of physical schemes is open and extensible. We have schemes for Eclipse projects, Java class resources, OSGI bundle resources, JDBC data sources, jar files, etc.

A \emph{logical} code location is akin to a fully qualified name. For each specific language we design a naming scheme for each source code element that is, in some sense, declared. An example of a logical location is \loc{java+class://myProject/java/util/List}. The scheme represents both the language and the kind of artifact that is identified. The authority declares the scope from which the name is resolved, in this case from \texttt{myProject} which depends on a particular version of the Java run-time. Finally, the path identifies the qualified name of the artifact in this scope. One goal of logical locations is to link uniquely to physical locations, at a certain moment in time, and at the same time be more or less stable under irrelevant code movement (such as moving the root source directory within a project). Another goal for such links is to be readable, writeable, recognizable and memorizable by human beings when developing new extractors, metrics or visuals. I.e. we might explore an \mthree\ model by projecting the information for an arbitrary class: the Rascal command \texttt{m@inheritance[|class://myPrj/java/util/List|]} would produce all interfaces that inherit from java.util.List.

The query part of a URI is used to \emph{modify} identities, for example to scope them for a version of a system: \loc{class://myPrj/java/util/List?svn=4242}. The offset and length fields are used to identify consecutive slice of characters of the identified artifact.

\mthree\ models are build on this concept of logical and physical source locations. It uses binary relations between locations, it annotates AST nodes with these locations and it embeds these locations into symbolic facts (such as types) to link back to source code whenever possible.

\subsection{Relations.} The \mthree\  model is both layered and compositional. This means that \mthree\ models can be combined (``linked'') and that they can be extended (``annotated''). The core relations are all between code locations: \emph{containment} defines which artifact is (logically) contained in which other artifact, \emph{declarations} define which logical locations are located at which physical locations, \emph{uses} defines which logical locations are used by which other logical locations.
An example containment tuple would be \loc{<|class:///foo/Bar|,|pkg:///foo|>}.

This core model is language independent, facilitating not only, volume metrics, browsing visuals (drill-down) and generic aggregation over containment relations, but also dependence between artifacts and thus impact and coupling/cohesion analyses. Also note that this core model is not restricted to handling programming languages. It can be used without doubt to model other kinds of formal languages like grammars, schema languages or even pictorial languages.

For modeling language specific information we annotate the above core model with extra relations. Again these are binary relations between logical locations. Examples for Java are \emph{inheritance}, \emph{overrides}, \emph{invocations}. These relations model key aspects of the static semantics of a programming language. Note that we never refer to instantiated or dynamic objects here, not even parametric type instantiations. All relations refer to source locations literally. For the accuracy of source code metrics, it is essential that \mthree\  separates what is written in the source code from what the code means dynamically. For example, if an abstract method from an interface is called we should not infer immediately all the call sites and add those to the invocations relation. Some metrics may want to count the fan-out to abstract methods, while other metrics want to know the impact on concrete implementations. You can compute this kind of information by composing basic facts, e.g. ``invocation $\circ$ overrides'' gives all the concrete callees for calls to abstract methods, and then compute a metric over the resulting relation instead.

\subsection{Trees.} For abstract syntax trees we use a general concept of algebraic data-types in Rascal. Every language comes with its own definitions. Algebraic data-types are easy to extend with new constructors (new programming language constructs). For \mthree\ we standardize some of the names used in defining AST types. In the core we standardize on five algebraic sorts to use when defining an abstract syntax: \texttt{Expression}, \texttt{Declaration}, \texttt{Statement}, \texttt{Type}, \texttt{Modifier}. The goal is to add as few as extra sorts as possible when adding a new language. This leads to models which \emph{over-approximate} the possible programs, but also increases the chance of reuse and extending existing fact extractors. For example, if all statements are in the same sort, then a basic function computing the cyclomatic complexity can be extended to cover a new language by just adding cases for the new types of statements (e.g. a \texttt{foreach} statement). We also provide annotations types for specific nodes, i.e. all nodes have a \texttt{src} annotation to point to the physical source location, all declarations may have a \texttt{decl} annotation to their logical location identifier and all Expressions may have a \texttt{type} annotation (see below).

Trees are useful mostly for the computation of metrics over code units that contain statements, such as cyclomatic complexity, but also to infer data and control flow information for use in the more advanced analyses. Trees are also expensive to keep in memory, so in M3 models they are always computed \emph{on-demand} for a particular logical location.

\subsection{Types} For types we introduce a single sort called \texttt{TypeSymbol}. We use this to represent any kind of abstract value that variables and expressions in a language may produce. For Java we have a default set of type symbols to represent (parametrized) class and interface types method signatures and its primitive types. These symbols can be used to compute with raw and parametrized types, either instantiated or uninstantiated. An example of a type symbol is: \loc{class(|class:///java/util/List, [class(|class:///java/lang/String|,[])])}, meaning the instantiated parametrized List type generated by the List class definition, and its type parameter is instantiated by the String class. We extended the core \mthree\  model with initial types: a relation from declarations to the types they generate and we annotate the trees of expressions with the types they produce. Using type symbols we may compute with and reason about dynamic artifacts that are never declared yet may exist at run-time. For example, a metrics for the number of possible instantiations of a parametrized type can be computed based on such information.

\subsection{Model composition} When we extract \mthree\  models we do this incrementally, i.e. per file, per project, per composition of a project with its dependencies. Each file (in a given programming language) produces one \mthree\  model. Then the models for all files in a project are fused into one single \mthree\  model by applying set union to all the relations of the model. Finally, if there are project dependencies, we may fuse the \mthree\  models for different projects. 

Some analyses are best done before fusion. We compute the volume of a project before we fuse in the declarations of the jars we depend on. Other analyses are done only after fusing: Depth of inheritance can only be computed if the models of classes we depend on our fully available. Since \mthree\  models are immutable values, like all Rascal values, it can never happen that we accidentally mix such models up. The \texttt{compose} function is called explicitly by the programmer to union the relation between two \mthree\ models and the \texttt{link} function does the same but updates the authority fields of all logical locations such that uses from one project may point to the declarations of another.

Currently we have extractors of \mthree models for jar files (i.e. from bytecode) from the JRE and Eclipse plugins, and from the source code of Eclipse project separately.  We then link these independently acquired M3 models to form complete models for further analysis. 

\section{Conclusion}

We have shown you a taste of \mthree, an extensible and composable model for source code artifacts based on relations and trees, with immutable value semantics, source location literals and extensible with annotations. It has support for basic language independent analyses and we have a detailed model for Java. Extensions to be expected soon are C\# and PHP support, and control flow and program dependence relations. We use \mthree\  in our course on Software Evolution at UvA and OU, and in the context of two research projects at CWI. At BENEVOL we hope to have discussion on its usability in a larger context of software analysis and software analytics.

\end{document}